\documentclass[11pt,a4paper,reqno]{amsart}%
\usepackage{amsthm,amsmath,amsfonts,amssymb,amsxtra,appendix,bookmark,dsfont,bm,mathrsfs,color}

\theoremstyle{plain}
\newtheorem*{theorem}{Theorem}

\theoremstyle{definition}

\newcounter{step} 

\usepackage{etoolbox}
\AtBeginEnvironment{proof}{\setcounter{step}{0}}

\DeclareMathOperator{\Tr}{Tr}

\def\bq{\begin{eqnarray}}
\def\eq{\end{eqnarray}}
\def\bqq{\begin{align*}}
\def\eqq{\end{align*}}

\def\nn{\nonumber}

\def\eps{\varepsilon}

\newcommand\1{{\ensuremath {\mathds 1} }}

\def\R {\mathbb{R}}
\def\C {\mathbb{C}}

\def\R {\mathbb{R}}
\def\C {\mathbb{C}}

\def\d{{\rm d}}

\title[] {Lieb-Thirring inequality with semiclassical constant and gradient error term}

\author[P. T. Nam]{Phan Th\`anh Nam}
\address{Department of Mathematics and Statistics, Masaryk University, Kotl\'a\v rsk\'a 2, 611 37 Brno, Czech Republic} 
\email{ptnam@math.muni.cz}

\begin{document}

\begin{abstract} In 1975, Lieb and Thirring derived a semiclassical lower bound on the kinetic energy for fermions, which agrees with the Thomas-Fermi  approximation up to a constant factor. Whenever the optimal constant in their bound coincides with the semiclassical one is a long-standing open question. We prove an improved bound with the semiclassical constant and a gradient error term which is of lower order.
\end{abstract}

\date{\today}

\maketitle

\setcounter{tocdepth}{1}

An important consequence of the Pauli exclusion principle is the fact that the kinetic energy of fermions increases much faster than the number of particles. In fact, a simple calculation shows that the lowest kinetic energy of $N$ identical fermions confined  in a unit volume in $\R^d$ is nearly $K^{\rm cl}_{d} N^{1+2/d}$ at leading order of $N$ large. Here 
\begin{equation} \label{eq:Ksc}
K^{\rm cl}_{d} := \frac{d}{d+2} \cdot \frac{4\pi^2}{(q\omega_d)^{2/d}}
\end{equation}
is the semiclassical constant, with $q$ the number of spin states ($q=2$ for electrons) and $\omega_d$ the volume of the unit ball in $\R^d$. In general, the semiclassical approximation of the kinetic energy of fermions in $\R^d$ is
\bq \label{eq:sc-app}
\Tr(-\Delta \gamma)  \approx K^{\rm cl}_{d}\int_{\R^d} \rho_\gamma(x)^{1+2/d} \d x
\eq
where $\gamma$ is the one-body density matrix of the system, which is  a trace class operator on $L^2(\R^d,\C^q)$ satisfying $0\le \gamma \le 1$, and $\rho_\gamma(x)=\Tr_{\C^q} \gamma(x,x)$ is the one-body density. The approximation \eqref{eq:sc-app} has been used widely in the Thomas-Fermi and related density-functional theories  \cite{Lieb-81}.

In 1975, Lieb and Thirring \cite{LieThi-75,LieThi-76} derived the rigorous lower bound 
\bq \label{eq:LT}
\Tr(-\Delta \gamma) \ge K_{d}\int_{\R^d} \rho_\Psi(x)^{1+2/d} \d x
\eq
for all one-body density matrices $\gamma$, with a universal constant $0<K_{d}<K_d^{\rm cl}$. Whenever \eqref{eq:LT} still holds true with $K_d$ replaced by $K^{\rm cl}_{d}$  (when $d\ge 3$) is a long-standing open question. Currently, the best justified constant for $d=3$ is $0.672\times K_d^{\rm cl}$ \cite{DolLapLos-08}. See also  \cite{Levitt-14} for a numerical investigation. 

In this note, we will derive an improved version of \eqref{eq:LT} with the semiclassical constant and an error term of lower order.  

\begin{theorem} If $\gamma$ is a trace class operator on $L^2(\R^d,\C^q)$ satisfying $0\le \gamma \le 1$, then for all $\eps>0$,
\bq \label{eq:main-ineq}
\Tr(-\Delta \gamma) \ge (1-\eps) K_d^{\rm cl} \int_{\R^d} \rho_\gamma(x)^{1+2/d} \d x - \frac{C_d}{\eps^{3+4/d} }  \int_{\R^d} \left|\nabla \sqrt{\rho_\gamma}(x)\right|^2 \d x.
\eq
Here $C_d>0$ is a universal constant depending only on the dimension. 
\end{theorem}

Note that our bound \eqref{eq:main-ineq} implies the Lieb-Thirring inequality \eqref{eq:LT} (with a non-sharp constant) by means of the Hoffmann-Ostenhof inequality \cite{HO-77} 
\begin{equation} \label{eq:HO}
\Tr(-\Delta \gamma)  \ge \int_{\R^d} \left|\nabla \sqrt{\rho_\gamma}(x)\right|^2 \d x.
\end{equation}

In most of physically interesting situations, the gradient term on the right side of \eqref{eq:HO} is much smaller than the kinetic energy. For example, at the ground state of $N$ ideal (i.e. non-interacting) fermions in a fixed volume, the gradient term  is proportional to $N$ while the kinetic energy grows as $N^{1+2/d}$ when $N$ becomes large. 

In principle, the gradient term corresponds to the uncertainty principle which is essentially a one-body effect, while the semiclassical term $\int \rho_\gamma^{1+2/d}$ captures the exclusion principle which is a truly many-body effect. The gradient   terms have also appeared in recent improvements of the Lieb-Oxford estimate on Coulomb exchange energy \cite{BenBleLos-12,LewLie-15}. 

When $d\le 2$, the optimal value of $K_{d}$ in \eqref{eq:LT} is expected to be  $K^{\rm So}_d< K_d^{\rm cl}$ where $K^{\rm So}_d$ is the best constant in Sobolev's inequality (a special case of \eqref{eq:LT} when $\gamma$ is a rank-one projection)
\bq \label{eq:So-GN}
\int_{\R^d} \left|\nabla \sqrt{\rho}(x)\right|^2 \d x \ge K^{\rm So}_d \int_{\R^d} \rho^{1+2/d} \d x, \quad \forall \rho\ge 0, \int \rho =1.
\eq
By contrast, our bound \eqref{eq:main-ineq} holds for all dimensions $d\ge 1$.

The original proof of the Lieb-Thirring inequality \eqref{eq:LT} in \cite{LieThi-75,LieThi-76} is based on its dual version
\begin{equation} \label{eq:LT-ev}
\Tr [-\Delta+V(x)]_- \ge - L_{d} \int_{\R^d} |V_-(x)|^{1+d/2} \d x
\end{equation}
with 
\bq \label{eq:LK}
L_d := \left( 1+\frac{d}{2}\right)^{-1} \left[ \left( 1+ \frac{2}{d}\right)K_d\right]^{-d/2}.
\eq
Here $a_-=\min(a,0)$ and the left side of \eqref{eq:LT-ev} is the sum of all negative eigenvalues of the Schr\"odinger operator $-\Delta+V$. The eigenvalue bound \eqref{eq:LT-ev} has its own interest and admits several generalizations; see e.g. \cite[Chapter 12]{LieLos-01} for further discussions. Unfortunately, this duality argument does not seem to apply to \eqref{eq:main-ineq}.

In 2011, Rumin found an elegant, direct proof of \eqref{eq:LT}. His ideas have been used to derive a positive density analogue of \eqref{eq:LT} in \cite{FraLewLieSei-13} and to provide a short proof of the Cwikel-Lieb-Rozenblum (CLR) bound in \cite{Frank-14}.
 
Recently, Lundholm and Solovej \cite{LunSol-13} found another direct proof of \eqref{eq:LT}. Their key observation is that \eqref{eq:LT} can be derived from a local version of the exclusion principle, which goes back to the first proof of the stability of matter by Dyson and Lenard \cite{DysLen-67} (the second proof of the stability of matter was given by Lieb and Thirring \cite{LieThi-75} by means of their inequality \eqref{eq:LT}). This approach has been used to prove Lieb-Thirring type inequalities for particles with fractional (anyon) statistics \cite{LunSol-13,LunSol-14}, for particles with point interactions \cite{FraSei-12} and for bosonic systems with homogeneous interactions \cite{LunPorSol-15,LunNamPor-16}.

Inspired by \cite{LunSol-13,FraSei-12,LunPorSol-15,LunNamPor-16}, we also use a localization argument to prove \eqref{eq:main-ineq}. Our new observation is that locally the density $\rho_\gamma(x)$ is close to a constant and the local contribution of the kinetic energy can be estimated by the sum of eigenvalues of Laplacian on a domain. This simplification allows us to recover the semiclassical constant, by means of a variant of the Berezin-Li-Yau inequality \cite{Berezin-72,LiYau-83}. The error of the localization procedure, which essentially comes from the uncertainty principle, can be controlled by the gradient term. The details are given below.

\medskip

\noindent
{\bf Proof of Theorem.} First, by a convexity argument, it suffices to consider the case $q=1$. We will denote by $C \ge 1$ a universal constant depending only on the dimension. The proof is divided into 3 steps.

\medskip

\noindent
{\bf Step 1.} By a density argument, it suffices to consider the case when $\rho_\gamma$ has compact support. We cover the support of $\rho_\gamma$ by a family of (finitely many) disjoint cubes $\{Q\}$, whose construction will be specified later. 

Since $\gamma$ is a trace class operator, it admits the spectral decomposition 
$$
\gamma= \sum_{i} \lambda _i |u_i\rangle \langle u_i|
$$
where $\{u_i\}$ is an orthonormal family in $L^2(\R^2)$ and $0\le \lambda_i \le 1$.  Then
\bq \label{eq:dec}
\Tr(-\Delta \gamma) = \sum_i \lambda_i \int_{\R^d} |\nabla u_i(x)|^2 \d x = \sum_Q \sum_i \lambda_i \int_Q |\nabla u_i(x)|^2 \d x .
\eq

Let us show that for every cube $Q$,
\bq \label{eq:local-LT}
\sum_i \lambda_i \int_Q |\nabla u_i(x)|^2 \d x \ge K_{d}^{\rm cl} |Q|^{-2/d}\left[\left(\int_Q \rho_\gamma \right)^{1+2/d} -C  \left(\int_Q \rho_\gamma \right)^{1+1/d}  \right].
\eq
If $\int_Q \rho_\gamma < 1$, then  \eqref{eq:local-LT} becomes trivial since the right side is negative with $C=1$. Therefore, we can assume $\int_Q \rho_\gamma \ge 1$. The left side of \eqref{eq:local-LT} can be rewritten as $\Tr[-\Delta_Q \gamma_Q]$, where $-\Delta_Q$ is the Neumann Laplacian on $Q$ and $\gamma_Q=\1_Q \gamma \1_Q$ (with $\1_Q$ the indicator/characteristic function of $Q$). Using $0\le \gamma_Q\le 1$ and explicit eigenvalues of the Neumann Laplacian on a cube, we can estimate
\begin{align*}
& \Tr[-\Delta_Q\gamma_Q] - |Q|^{-2/d} \mu \int_Q \rho_\gamma = \Tr[(-\Delta_Q-|Q|^{-2/d}\mu)\gamma_Q] \\
&\ge  \Tr[-\Delta_Q-|Q|^{-2/d} \mu]_-= |Q|^{-2/d} \sum_{p\in (\mathbb{N}_0)^d} \Big[\pi^2 |p|^2-\mu\Big]_-\\
&=|Q|^{-2/d} \sum_{k=0}^d  {d \choose k} \sum_{r\in \mathbb{N}^k} [\pi^2 |r|^2-\mu]_- 
\end{align*}
for all $\mu>0$. Here recall that $a_-=\min(a,0)$ and we distinguished $\mathbb{N}_0=\{0,1,2...\}$ from $\mathbb{N}=\{1,2,...\}$. By the Berezin-Li-Yau inequality \cite{Berezin-72,LiYau-83} for the sum of Dirichlet eigenvalues (with $L^{\rm cl}_d$ being related to $K^{\rm cl}_d$ as in \eqref{eq:LK})
$$ 
\sum_{r\in \mathbb{N}^k} \Big[\pi^2 |r|^2-\mu\Big]_- \ge - L^{\rm cl}_k \mu^{1+k/2},
$$
we obtain
\begin{align*}
\Tr[-\Delta_Q\gamma_Q]  \ge  |Q|^{-2/d} \left[ \mu \int_Q \rho_\gamma -  L_d^{\rm cl} \mu^{1+d/2} - C \sum_{k=0}^{d-1}  \mu^{1+k/2}\right]
\end{align*}
for all $\mu>0$. Optimizing over $\mu$ leads to \eqref{eq:local-LT} (the contribution of $\mu^{1+k/2}$ with $k<d-1$ is small because $\int_Q \rho_\gamma \ge 1$). 

Note that \eqref{eq:local-LT} is essentially equivalent to a Weyl asymptotic estimate on the sum of Neumann eigenvalues on a cube; see \cite{Kroger-94} for related results. 

From \eqref{eq:dec} and \eqref{eq:local-LT}, it follows that 
\bq \label{eq:lwb-1}
\Tr(-\Delta \gamma) \ge  K_{d}^{\rm cl}  \sum_Q  |Q|^{-2/d}\left[\left(\int_Q \rho_\gamma \right)^{1+2/d} -C  \left(\int_Q \rho_\gamma \right)^{1+1/d}  \right].
\eq

\noindent
{\bf Step 2.} We show that for every $0<\Lambda \le \Tr \gamma$, it is possible to choose the cubes $\{Q\}$ such that 
\bq \label{eq:cv-1}
\int_Q \rho_\gamma\le \Lambda, \quad \forall Q,
\eq 
and
\bq \label{eq:cv-2}
 \sum_Q  |Q|^{-2/d}\left[C\Lambda^{-1/d}\left(\int_Q \rho_\gamma \right)^{1+2/d} -  \left(\int_Q \rho_\gamma \right)^{1+1/d}  \right] \ge 0.
\eq 

The construction of the cubes is similar to \cite[Lemma 9]{LunNamPor-16}: first we cover the support of $\rho_\gamma$ by a big cube, then divide it into $2^d$ sub-cubes of half side length; if the integral of $\rho_\gamma$ on any sub-cube is still bigger than $\Lambda$, then we continue dividing this sub-cube into $2^d$ (sub-) sub-cubes of half side length and repeat the procedure. After finitely many steps, we obtain a collection of disjoint sub-cubes $\{Q\}$  satisfying \eqref{eq:cv-1}. Moreover, we can distribute the sub-cube $Q$ into disjoint groups such that in each group:

(i) The integral of $\rho_\gamma$ over the union of the smallest sub-cubes is $\ge \Lambda$;

(ii) There are at most $2^d$ sub-cubes of every given size.

Consider an arbitrary group $\mathcal{G}$. Let $m$ be the volume of the smallest sub-cubes in $\mathcal{G}$. Then the volumes of all sub-cubes in $\mathcal{G}$ are $m2^{dk}$ with $k=0,1,2,...$ and there are at most $2^d$ sub-cubes for each $k$. We have
\begin{align*}
\max_{\substack{Q\in \mathcal{G}\\|Q|=m}}  |Q|^{-2/d} \left(\int_Q \rho_\gamma \right)^{1+2/d} &\ge m^{-2/d} \left(\frac{\Lambda}{2^d} \right)^{1+2/d},\\
\sum_{Q\in \mathcal{G}}  |Q|^{-2/d} \left(\int_Q \rho_\gamma \right)^{1+1/d} & \le\sum_{k=0}^\infty \sum_{\substack{Q\in \mathcal{G}\\ |Q|=m2^k}} (m2^{dk})^{-2/d} \Lambda^{1+1/d}\\
&\le \sum_{k=0}^\infty 2^d (m2^{dk})^{-2/d} \Lambda^{1+1/d} = \frac{4}{3} \cdot 2^d m^{-2/d} \Lambda^{1+1/d}. 
\end{align*}
Thus, with $C=4^{d+2}/3$, 
$$
 \sum_{Q\in \mathcal{G}}  |Q|^{-2/d}\left[C\Lambda^{-1/d}\left(\int_Q \rho_\gamma \right)^{1+2/d} -  \left(\int_Q \rho_\gamma \right)^{1+1/d}  \right] \ge 0.
$$
Summing over all groups leads to \eqref{eq:cv-2}. From  \eqref{eq:lwb-1} and \eqref{eq:cv-2}, it follows that 
\bq \label{eq:K-01}
\Tr(-\Delta \gamma) \ge  K_{d}^{\rm cl} (1-C\Lambda^{-1/d}) \sum_Q  |Q|^{-2/d}\left(\int_Q \rho_\gamma \right)^{1+2/d}.
\eq

\noindent
{\bf Step 3.} To conclude, let us show that 
\begin{align} \label{eq:PS-rho} |Q|^{-2/d} \left( \int_{Q} \rho_\gamma \right)^{1+2/d} &\ge (1+\eps)^{-(1+4/d)} \int_Q \rho_\gamma^{1+2/d} \nn\\
&\quad - C\eps^{-(1+4/d)} \left( \int_Q |\nabla \sqrt{\rho_\gamma}|^2 \right) \left(\int_Q \rho_\gamma \right)^{2/d}
\end{align}
for every $\eps>0$. Denote 
$$u(x)=\sqrt{\rho_\gamma}(x), \quad u_Q=  |Q|^{-1}\int_{Q} u(x) \d x.$$
For all $\eps>0$, by H\"older's inequality 
$$
\left(1 + \eps \right)^{p-1} \left(|u(x)-u_Q|^p + \eps^{1-p}|u_Q|^p \right) \ge \left( |u(x)-u_Q|+ |u_Q| \right)^p  \ge |u(x)|^{p}
$$
with $p=2+4/d$, we can write
$$
|u_Q|^{2+4/d} \ge (1+\eps)^{-(1+4/d)} |u(x)|^{2+4/d} - C \eps^{-(1+4/d)} |u(x)-u_Q|^{2+4/d}.
$$
Next, integrate the latter estimate over $x\in Q$, then use H\"older's inequality
$$
\int_{Q} |u|^2 \ge |Q|^{-1} \left( \int_Q |u| \right)^{2} = |Q| |u_Q|^2
$$
for the left side and the Poincar\'e-Sobolev inequality \cite[Theorems 8.12]{LieLos-01}
\begin{align*}
\int_Q \left| u(x) -u_Q\right|^{2+4/d} \d x \le C \left( \int_Q |\nabla u(x)|^2 \d x  \right) \left(\int_Q |u(x)|^2 \d x\right)^{2/d}
\end{align*}
for the right side. All this leads to \eqref{eq:PS-rho} as $u=\sqrt{\rho_\gamma}$. 

Finally, insert \eqref{eq:PS-rho} into \eqref{eq:K-01} and use \eqref{eq:cv-1}. We obtain
\begin{align*}
\Tr (-\Delta \gamma ) \ge K_{d}^{\rm cl} (1-C\Lambda^{-1/d}) \left[ (1+\eps)^{-(1+4/d)}  \int_{\R^d} \rho_\gamma^{1+2/d} \right. \qquad\qquad\\ 
\left. \qquad - C  \eps^{-(1+4/d)} \Lambda^{2/d} \int_{\R^d} |\nabla \sqrt{\rho_\gamma}|^2 \right]
\end{align*}
for all $\eps>0$ and $0<\Lambda\le \Tr \gamma$. Choosing $\eps=\Lambda^{-1/d}$, we get
\bq \label{eq:K-02}
\Tr(-\Delta \gamma) \ge K_d^{\rm cl} (1-C\eps) \int_{\R^d}\rho_\gamma^{1+2/d} - C \eps^{-(3+4/d)} \int_{\R^d} |\nabla \sqrt{\rho_\gamma}|^2
\eq
for all $\eps \ge (\Tr \gamma)^{-1/d}$. 

If $\eps>1$, then $1-C\eps<0$ and the right side of \eqref{eq:K-02} is negative. If $\eps<\min( (\Tr \gamma)^{-1/d},1)$, then the right side of \eqref{eq:K-02} is negative when $C K_d^{\rm So}>1$ because of \eqref{eq:So-GN}. Therefore, \eqref{eq:K-02} holds true for all $\eps>0$. It is equivalent to \eqref{eq:main-ineq} and the proof is complete. \qed

Finally, let us remark that our proof of \eqref{eq:main-ineq} is very general and it could be generalized in many other situations as soon as one has a Weyl asymptotic estimate on cubes.

{\bf Acknowledgement.} I thank Simon Larson and Douglas Lundholm for helpful comments.


\begin{thebibliography}{11}


\bibitem{BenBleLos-12} R. D. Benguria, G. A. Bley, and M. Loss, A new estimate on the indirect Coulomb energy, Int. J. Quantum
Chem. 112 (2012), 1579.

\bibitem{Berezin-72} F. A. Berezin, Covariant and contravariant symbols of operators, Izv. Akad. Nauk SSSR Ser. Mat.
13 (1972),  1134-1167.


\bibitem{DolLapLos-08} J. Dolbeault, A. Laptev, and M. Loss, Lieb-Thirring inequalities with improved constants,
J. Eur. Math. Soc. 10 (2008),  1121-1126.

\bibitem{DysLen-67} F. J. Dyson and A. Lenard, Stability of matter. I, J. Math. Phys. 8 (1967),  423-
434; II. J. Math. Phys. 9 (1968),  698-711.

\bibitem{Frank-14} R. L. Frank, Cwikel's theorem and the CLR inequality. J. Spectral Theory 4 (2014), no. 1, 1-21.

\bibitem{FraLewLieSei-13} R. L. Frank, M. Lewin, E.H. Lieb and R. Seiringer, A positive density analogue of the Lieb-Thirring inequality. Duke Math. J. 162 (2013), no. 3, 435-495

\bibitem{FraSei-12} R. L. Frank and R. Seiringer, Lieb-Thirring inequality for a model of particles
with point interactions, J. Math. Phys., 53 (2012),  095201.

\bibitem{Kroger-94} P. Kroger, Estimates for sums of eigenvalues of the Laplacian, J. Funct. Anal. 126 (1994), 217-227.


\bibitem{LewLie-15} M. Lewin and E. H. Lieb, Improved Lieb-Oxford exchange-correlation inequality with gradient correction. Phys. Rev. A 91 (2015), 022507. 

\bibitem{Levitt-14} A. Levitt, Best constants in Lieb-Thirring inequalities: a numerical investigation. Journal of Spectral Theory 4.1 (2014),  153-175

\bibitem{LiYau-83} P. Li and S.T. Yau, On the Schr\"odinger equation and the eigenvalue problem. Commun. Math.
Phys. 88 (1983),  309-318.

\bibitem{Lieb-81} E. H. Lieb, Thomas-Fermi and related theories of atoms and molecules, Rev. Mod. Phys. 53 (1981),  603-641.

\bibitem{LieLos-01} E.~H. Lieb and M.~Loss, Analysis, Second edition, Graduate Studies in Mathematics, American Mathematical Society, Providence, RI, 2001.
	
\bibitem{LieThi-75} E. H. Lieb and W. E. Thirring, Bound on kinetic energy of fermions which proves
stability of matter, Phys. Rev. Lett. 35 (1975),  687-689.

\bibitem{LieThi-76} E. H. Lieb and W. E. Thirring, Inequalities for the moments of the eigenvalues of the Schr\"odinger Hamiltonian and their relation to Sobolev inequalities, in Studies in Mathematical Physics,
Princeton University Press, 1976,  269-303.

\bibitem{LunNamPor-16} D. Lundholm,  P. T. Nam, and F. Portmann, Fractional Hardy-Lieb-Thirring and related inequalities for interacting systems. Arch. Rational Mech. Anal. 219 (2016), 1343-1382.

\bibitem{LunPorSol-15} D. Lundholm, F. Portmann, and J. P. Solovej, Lieb-Thirring bounds for interacting
Bose gases, Commun. Math. Phys. 335 (2015),  1019-1056.


\bibitem{LunSol-13} D. Lundholm and J. P. Solovej, Hardy and Lieb-Thirring inequalities for anyons,
Commun. Math. Phys. 322 (2013),  883-908.

\bibitem{LunSol-14} D. Lundholm and J. P. Solovej, Local exclusion and Lieb-Thirring inequalities for intermediate and fractional statistics, Ann. Henri Poincar\'e 15 (2014), 1061-1107.


\bibitem{HO-77} M. Hoffmann-Ostenhof and T. Hoffmann-Ostenhof, Schr\"odinger inequalities
and asymptotic behavior of the electron density of atoms and molecules, Phys. Rev.
A 16 (1977),  1782-1785.

\bibitem{Rumin-11} A. Rumin, Balanced distribution-energy inequalities and related entropy bounds, Duke
Math. J., 160 (2011), 567-597.

\end{thebibliography}
\end{document}